\documentclass[14pt,a4paper]{extarticle}

\usepackage[utf8]{inputenc}
\usepackage[english]{babel}
\usepackage[T2A]{fontenc}

\usepackage{amsmath}
\usepackage{amsfonts}
\usepackage{amssymb}
\usepackage{graphicx}
\usepackage{hyperref}
\usepackage{color}
\usepackage{fullpage}
\usepackage{indentfirst}
\usepackage{tabularx}
\usepackage{cite}
\usepackage{diagbox}

\definecolor{darkblue}{RGB}{23,26,73}
\definecolor{darkgreen}{RGB}{27,37,14}
\definecolor{darkred}{RGB}{37,14,14}

\hypersetup{%
  colorlinks=true,%
  linkcolor=darkred,
  citecolor=darkgreen,
  urlcolor=darkblue,
  pdfauthor={A. A. Lagutin},%
  pdftitle={Diffusive shock acceleration: non-classical model of cosmic ray transport},%
  pdfsubject={Manuscript of an article for the journal PHAN},%
  pdfkeywords={Cosmic rays, diffusive shock acceleration, non-classical particle transport, L\'{e}vy flights, L\'{e}vy  traps, energy spectral index},%
  breaklinks=true,%
  plainpages=false%
}

\begin{document}

\begin{center}
{\large\bf Diffusive shock acceleration: non-classical model of cosmic ray transport}

\vspace*{1cm}

A. A. Lagutin\footnote{\url{https://orcid.org/0000-0002-1814-8041}}

\vspace*{5mm}

Altai State University, Barnaul, 656049 Russia

\vspace*{5mm}

e-mail: \verb|lagutin@theory.asu.ru|

\end{center}

\begin{abstract}

In this work the theory of diffusive shock acceleration is extended to the case of non-classical particle  transport with L\'{e}vy flights and L\'{e}vy  traps, when the mean square displacement grows nonlinearly with time. In this approach the Green function is not a Gaussian but it exhibits power-law tails. By using the propagator appropriate for non-classical diffusion, it is found for the first time that energy spectral index of particles accelerated at shock front is $\gamma = [\alpha (\mathrm{r} + 5) -  6 \beta]/[\alpha(\mathrm{r}-1)]$, where 
$0 < \alpha < 2$ and $0 <\beta < 1$ are the exponents of power-law behavior of L\'{e}vy flights and L\'{e}vy traps, respectively.
We note that this result coincides with standard slope at $\alpha=2, \beta=1$ (normal diffusion),  and also includes those obtained earlier for the subdiffusion ($\alpha=2, \beta<1$) and superdiffusion ($\alpha<2,  \beta=1$) regimes.

\vspace*{5mm}

\textbf{Keywords:} cosmic rays, diffusive shock acceleration, non-classical particle transport, L\'{e}vy flights, L\'{e}vy traps, energy spectral index.

\end{abstract}

\section*{Introduction}

Galactic cosmic rays up to about 100 PeV are believed to be accelerated by shock waves at supernova remnants by a Fermi process called diffusive shock acceleration (DSA)~\cite{Ptuskin:2003}. This process is based on the assumption that cosmic ray (CR) particles can be confined near the shock discontinuity by encounters with both inhomogeneities in the interstellar space and with shock. The scattering mean-free path is assumed to be much larger than the shock thickness and much shorter than the length of the area where the wave activity upstream and downstream is strong. As a result particles can cross the shock discontinuity repeatedly. The charged particles gain energy through the repeated scatterings off the converging up- and downstream scattering centers.

Understanding this particle acceleration mechanisms occurred in 1977-1978. 
In pioneering articles~\cite{Krymskii:1977,Blandford:1978,Axford:1977,Bell:1978} the authors independently showed how a power-law momentum spectrum of accelerated test particles results from very general properties of a plasma shock.

The calculations were carried out using two different approaches under the assumption that the particle transport in the acceleration zone is described by the normal diffusion model, in which the mean square displacement is proportional to time. In the first approach~\cite{Krymskii:1977,Blandford:1978,Axford:1977}, the spectrum of accelerated particles was found by solving the simple one-dimensional diffusion-convection equation with a flow discontinuity representing the shock transition. Bell~\cite{Bell:1978}, on the other hand, employed an equivalent individual particle kinetic approach to obtain the distribution function. In this microscopic description superthermal particles are assumed to scatter elastically in the local plasma frame and to have giroradii much longer than the shock thickness. Particles freely scatter from one side of the shock to the other and scatter against converging plasma, increasing energy every time they cross the shock.

Both approaches showed that for highly relativistic particles, DSA gives rise to a power-law energy distribution with spectral index $\gamma = (\mathrm{r}  + 2)/(\mathrm{r}  - 1)$, where $\mathrm{r} $ is the compression ratio of the shock, i.e. the ratio between the unshocked plasma speed $u_1$ and the shocked plasma speed $u_2$ measured in the shock frame. The cosmic rays spectral index for a maximal compression ratio of four equals to $\gamma = 2$.

From the very beginning it was understood that the spectrum of test particles accelerated at a shock is 
independent upon the details of particle scattering (for instance diffusion coefficient) and is close to what is 
required by the observed CR spectrum at the Earth, after accounting for transport effects.

Numerous studies of the character of the distribution of matter and magnetic fields in the interstellar medium and planetary environments, carried out over the past few decades, have made it possible to find much evidence of the existence of structures that differ not only in their scale, but also in their morphology~\cite{Elmegreen:1996,Heyer:1998,Elmegreen:2004,Bergin:2007,Sanchez:2008,Efremov:2003,Fuente:2009,Sanchez:2010,Shi:2025,Raptis:2025}. It has been established that structures such as filaments, ribbons, clouds, voids and transient systems are widespread formations in the interstellar medium. The rich diversity of structures may be related to a fundamental property of turbulence called intermittency~\cite{Zeldovich:1990}. The fluctuating (or small-scale) turbulent dynamo mechanism, i.e., a random flow of electrically conducting fluid that generates a random magnetic field with zero mean~\cite{Zeldovich:1990,Wilkin:2007}, and random shock waves~\cite{Bykov:1987} create highly intermittent magnetic fields with random magnetic structures surrounded by weaker fluctuations. It was also shown that the propagation of cosmic rays in such an interstellar medium depends on magnetic intermittency in the energy range $E\lesssim 10^9$~GeV~\cite{Shukurov:2017}.

Non-homogeneous character of matter distribution and associated magnetic field noted above should be adequately incorporated into the cosmic ray diffusion model of particle transport in the vicinity of a shock front. 
The need to change the standard scenario of acceleration and propagation of cosmic rays is also due to recent results of
experiments~\cite{dampe-p:2019,dampe-he:2021,calet:2023,calet-p:2022,calet-he:2023,lhaaso:2025all,lhaaso:2025p}. For example, in our work~\cite{lagutin-bras:2025} shows that the power-law behavior of the the spectra of protons and helium nuclei before and after the break around 10$-$30~TeV from a group of tevatrons 
and the soft spectrum of particle generation in these sources contradict the standard scenario.

The main goal of this work is to extend the theory of diffusive shock acceleration to the case of non-classical particle transport with
L\'{e}vy flights and L\'{e}vy traps, when the mean square displacement grows nonlinearly with time.

\section{Non-classical diffusion model}

In the papers~\cite{Lagutin:2001CRC,Lagutin:2001NP,Lagutin:2003}, for the first time, a generalization of the model of normal CRs diffusion~\cite{Ginzburg:1964} was performed for the case of propagation in the interstellar medium, which has the properties described above. The authors of these works proposed an original approach, which consists of replacing the assumption of statistical homogeneity of the distribution of inhomogeneities with a more general statement about the fractal nature of their distribution. An important consequence of this assumption is the power-law distribution of the free paths of particles $r$ between inhomogeneities $p(r) \propto A(E,\alpha) r^{-\alpha - 1}, r \rightarrow \infty, 0 < \alpha < 2$ (L\'{e}vy flights), as well as the power-law distribution of the time $t$ that particles spend in them (L\'{e}vy traps) $q(t) \propto B(E,\beta)t^{-\beta - 1}, t \rightarrow \infty, \beta < 1$.

Taking into account the power-law asymptotics of the distributions $p(r)$ and $q(t)$ in the procedure for deriving the diffusion equation within the framework of the semi-analytical Motroll$\&$Weiss model~\cite{Montroll:1965} (or the continuous time random walk), which is widely used in this class of problems, leads to a generalized non-classical diffusion equation~\cite{Lagutin:2001CRC,Lagutin:2001NP,Lagutin:2003} with fractional differential operators~--- the fractional Laplacian $(-\Delta)^{\alpha/2}$ (the Riesz operator)~\cite{Samko:1993} and the fractional Riemann-Liouville derivative $\mathrm{D}_{0+}^{\beta}$~\cite{Samko:1993}~--- reflecting the non-locality and non-Markovian nature of the diffusion process, respectively.

In the one-dimensional case discussed in this paper, it is easy to write a one-dimensional version of the generalized equation of non-classical diffusion. The equation for the particle density with energy $E$ at time $t$ and a distance $x$ from the source with density $Q(x,t,E)$ without taking into account energy losses and nuclear interactions of cosmic rays is written in the form
\begin{equation}\label{eq:nonclass-diff-eq}
\frac{\partial N(x, t, E)}{\partial t}= -D(E, \alpha, \beta)\mathrm{D}_{0+}^{1-\beta}\left(-\dfrac{\partial^2}{\partial x^2}\right)^{\alpha/2} N(x, t, E) + Q(x, t, E).
\end{equation}
Here $D(E, \alpha, \beta) = D_0(\alpha, \beta) \left(E/1~\text{GeV}\right)^{\delta}$ is the anomalous diffusivity. The operator $(-\partial^2/\partial x^2)^{\alpha/2}$ is a one-dimensional fractional Laplacian, $\mathrm{D}_{0+}^{\beta}$ is the Riemann-Liouville fractional derivative.   Note that for $\alpha=2$, $\beta=1$ from~\eqref{eq:nonclass-diff-eq} we obtain the Ginzburg-Syrovatskii normal diffusion equation~\cite{Ginzburg:1964} in the one-dimensional case.
  
The equation for the Green's function $G(x, t, E; x_0, t_0, E_0)$, which determines the probability of finding a particle at a point $x$ of the medium at time $t$ in a unit energy interval around $E$, if the source emitted one particle with energy $E_0$ from point $x_0$ at time $t_0$, we write in the form
\begin{multline}\label{eq:green-func}
\frac{\partial G(x, t, E; x_0, t_0, E_0)}{\partial t}=-D(E, \alpha, \beta)\mathrm{D}_{0+}^{1-\beta}\left(-\dfrac{\partial^2}{\partial x^2}\right)^{\alpha/2} G(x, t, E; x_0, t_0, E_0) +\\ 
+\delta(x-x_0)\delta(t-t_0)\delta(E-E_0).
\end{multline}
The solution to the equation~\eqref{eq:green-func} is found using the Fourier-Laplace transforms in spatial and temporal coordinates. The Green's function in this case is equal to
\begin{multline}\label{eq:green-function}
G(x, t, E; x_0, t_0, E_0) = \left[D(E,\alpha,\beta)(t - t_0)^{\beta}\right]^{-1/\alpha}\times\\
\times\Psi_1^{(\alpha,\beta)}\left[\left(x - x_0\right) \left(D(E,\alpha,\beta)(t - t_0)^{\beta}\right)^{-1/\alpha}\right] \delta(E-E_0).
\end{multline}
Here
\begin{equation*}\label{eq:psi-stable-distribution}
\Psi_1^{(\alpha, \beta)}(r)=\int\limits_0^\infty g_1^{(\alpha)}\left(r\tau^{\beta/\alpha}\right) q_1^{(\beta)}(\tau) \tau^{\beta/\alpha} d\tau
\end{equation*}
is the density of the fractional-stable distribution~\cite{Uchaikin:1999a}, determined by the one-dimensional symmetric stable distribution $g_1^{(\alpha)}(r)$ ($0<\alpha\leqslant 2$) and the one-sided stable distribution $q_1^{(\beta)}(t)$ with the characteristic parameter $0<\beta\leqslant 1$. The mathematical description of these distributions is presented in the Appendix. 

Note that for $\alpha=2$ and $\beta=1$ the density $\Psi_1^{(\alpha, \beta)}(r)$ is the Gaussian normal distribution and from~\eqref{eq:green-function} we obtain the well-known result of the normal diffusion model~\cite{Ginzburg:1964}.

The Green's function~\eqref{eq:green-function} allows us to find a solution to the non-classical diffusion equation~\eqref{eq:nonclass-diff-eq} for the cosmic ray particle density $N(x, t, E)$ in the case of sources described by the density $Q(x, t, E)$. By definition, we have
\begin{equation}\label{eq:NGS}
N(x,t,E) = \int\limits_{\mathrm{R}^1} dx_0 \int\limits_{-\infty}^t d t_0 \int\limits_{E}^{\infty} d E_0  G(x-x_0,t-t_0,E-E_0)Q(x_0,t_0,E_0).
\end{equation}

\section{DSA: The spectral index in non-classical\\ diffusion model}

In this paper, the calculation of the spectral index of accelerated particles in non-classical model of cosmic ray transport is carried out within the framework of the approach proposed in~\cite{Bell:1978}. The author~\cite{Bell:1978} calculates the mean energy gain each time a cosmic ray particle crosses the shock and derives the energy spectrum by balancing the energy gain against the probability of a CR escaping downstream. It was shown that the differential spectrum of accelerated particles has a power-law energy distribution with spectral exponent $\gamma$ = (r + 2)/(r - 1) under the assumption that the particle transport in the acceleration zone is described by the normal diffusion model~\cite{Ginzburg:1964}.

In a more general form, by definition (see paragraph 12.2 in~\cite{Gaisser:2016}), this index has the form
\begin{equation}\label{N21}
\gamma = - \ln (1 - P_{\text{esc}})/\ln(1 + \xi) + 1 \approx P_{\text{esc}}/\xi  +1,   
\end{equation}
where $P_{esc}$ is the probability for a particle to escape from the acceleration region, $\xi = \Delta E/E = (4/3)(u_1 - u_2)/V$ is the relative gain of energy for particles. 

It should be noted that the key result of Bell's approach~\cite{Bell:1978} --- the integral energy spectral index of accelerated particles $\tilde {\gamma} = - \ln (1 - P_{\text{esc}})/\ln(1 + \xi)$ --- was experimentally verified in a recent paper~\cite{Barontini:2025}.
It has been shown that the index  observed in experiment is remarkably compatible with the functional dependence as predicted by Bell's argument.

In~\cite{Bell:1978} the escape probability is given by the ratio of the particle flux $\Phi_2$ exiting from downstream over 
the incident flux $\Phi_1$ coming from upstream: $P_{\text{esc}} = \Phi_2/\Phi_1$. With the assumption of isotropic particle distribution, the flux of particles crossing the shock from upstream to downstream is $\Phi_1 = N_0 V/4$~\cite{Bell:1978}, where
$N_0$ is the density of particles with velocity $V$ at the shock in the shock frame. The particle flux exiting from downstream is $\Phi_2 = N_2 u_2$, where $N_2$ is the far downstream density. In the work~\cite{Kirk:1996} it was shown that in the stationary model the particle density in the far downstream, regardless of the diffusion regime,  is equal to $N_2=Q_0/V_{\text{sh}}$, where $Q_0$ is the injection rate of particles by the shock, and $V_{\text{sh}}$ is its speed.

Using  $\Phi_1$ and $\Phi_2$, we can represent the escape probability as
\begin{equation}\label{N22}
P_{\text{esc}} = 4 Q_0 u_2/N_0 V V_{\text{sh}}.   
\end{equation}
To determine the density of particles at the shock $N_0$, the Green's function~\eqref{eq:green-function} and equality~\eqref{eq:NGS} are used.
Considering that the source of particles is a shock front moving at a speed $V_{\text{sh}}$, i.e. the density of source is equal to 
\begin{equation*}
Q(x_0,t_0,E_0)=Q_0(E_0)\delta(x_0 - V_{\text{sh}}t_0),
\end{equation*}
where $Q_0$ is the injection rate of particles by the shock, we find the density of particles at point $x$ at time $t$ with energy $E$:
\begin{multline}\label{P_{esc}}
N(x,t,E) = \int\limits_{\mathrm{R}^1} dx_0 \int\limits_{-\infty}^t d t_0 \int\limits_{E}^{\infty} d E_0 Q_0(E_0) \left[D(E,\alpha,\beta)(t - t_0)^{\beta}\right]^{-1/\alpha}\times\\
\times\Psi_1^{(\alpha,\beta)}\left[\left(x - x_0\right) \left(D(E,\alpha,\beta)(t - t_0)^{\beta}\right)^{-1/\alpha}\right] \delta(E-E_0) \delta(x_0 - V_{\text{sh}}t_0).
\end{multline}
Integrating in equation~\eqref{P_{esc}} over $x_0$ and $E_0$, we obtain
\begin{multline}\label{N31}
N(x,t,E) = Q_0(E)\int\limits_{-\infty}^t d t_0 \left[D(E,\alpha,\beta)(t - t_0)^{\beta}\right]^{-1/\alpha}\times\\
\times\Psi_1^{(\alpha,\beta)}\left[\left(x - V_{\text{sh}}t_0\right) \left(D(E,\alpha,\beta)(t - t_0)^{\beta}\right)^{-1/\alpha}\right]. 
\end{multline}
By introducing a new variable $\tau = t - t_0$, after calculations we obtain an equation that allows us to find $N_0$:
\begin{multline}\label{N32}
N(x,t,E) = Q_0(E)\int\limits_0^{\infty} d \tau \left[D(E,\alpha,\beta)\tau^{\beta}\right]^{-1/\alpha}\times\\\times\Psi_1^{(\alpha,\beta)}\left[\left(x - V_{\text{sh}}t + V_{\text{sh}}\tau\right) \left(D(E,\alpha,\beta)\tau^{\beta}\right)^{-1/\alpha}\right]. 
\end{multline}
Indeed, since we are interested in the density of particles at the shock wave front, i.e. at the point $x = V_{\text{sh}}t$, 
then $ N(V_{\text{sh}}t,t,E)\equiv N_0$. Thus, assuming in equation~\eqref{N32} $x = V_{\text{sh}}t$, we find
\begin{equation}\label{N4}
N_0 = Q_0(E) \int\limits_{0}^{\infty}\frac{d\tau}{(D(E,\alpha,\beta)\tau^{\beta})^{1/\alpha}} \Psi_1^{(\alpha,\beta)}\left( \frac{V_{\text{sh}} \tau}{(D(E,\alpha,\beta)\tau^{\beta})^{1/\alpha}}\right).
\end{equation}
To calculate the integral in~\eqref{N4} the new variable $\rho= {V_{\text{sh}} \tau}/{(D(E,\alpha,\beta)\tau^{\beta})^{1/\alpha}}$ is used. Considering that $d\rho$ is given by the equality
\begin{equation*}\label{N41}
d\rho = \left(V_{\text{sh}}/(D(E,\alpha,\beta)^{1/\alpha}\right) (1 - \beta/\alpha) \tau^{- \beta/\alpha}d\tau, 
\end{equation*}
we have:
\begin{equation}\label{N42}
\frac{d\tau}{D(E,\alpha,\beta)^{1/\alpha}\tau^{\beta/\alpha}} =\frac{d\rho}{(1 - \beta/\alpha)V_{\text{sh}}}.
\end{equation}
Substituting~\eqref{N42} into equation~\eqref{N4}, we find
\begin{equation}\label{N43}
N_0 = \frac{Q_0(E)}{(1 - \beta/\alpha)V_{\text{sh}}} \int\limits_{0}^{\infty} d\rho \Psi_1^{(\alpha,\beta)}(\rho).
\end{equation}
Since 
$$
\int\limits_{0}^{\infty} d\rho \Psi_1^{(\alpha,\beta)}(\rho)= 1/2,
$$
we finally obtain
\begin{equation}\label{N5}
N_0(E) = \frac {Q_0(E)}{2(1 - \beta/\alpha)V_{\text{sh}}}.
\end{equation}
It should be noted that in the case $\alpha=2$, $\beta= 1$ from~\eqref{N5} we get classical result $N_0=Q/V_{\text{sh}}$.

The result obtained above allows us to find the escape probability. From~\eqref{N22} we have:
\begin{equation}\label{N6}
P_{\text{esc}} = 8(1 - \beta/\alpha) u_2/V.
\end{equation}
In case of normal diffusion ($\alpha=2$, $\beta= 1$) from~\eqref{N6} we find classical result~\cite{Bell:1978}: $P_{\text{esc}} = 4 u_2/V$.

Substituting the escape probability $P_{\text{esc}}$ and the relative gain of energy for particles $\xi$ into equation~\eqref{N21} we obtain the spectral index of accelerated particles in non-classical model of cosmic ray transport:
\begin{equation*}\label{N7}
\gamma = \frac{\alpha(\mathrm{r} + 5) - 6 \beta}{\alpha(\mathrm{r} -1)}.
\end{equation*}

The numerical values of the spectral index for two compression ratios $\mathrm{r}=4$ and $\mathrm{r}=3$ of the shock  are given in the Tables~\ref{tab:r4}, \ref{tab:r3}. The right column in the tables describes the subdiffusion regime, and the bottom row gives the index values for the superdiffusion regime.

\begin{center}
\begin{table}[p]
\centering
\caption{The spectral index $\gamma$ for a shock with compression ratio $\mathrm{r}=4$}\label{tab:r4}
\begin{tabular}{|w{c}{2cm}|w{c}{1cm}|w{c}{1cm}|w{c}{1cm}|w{c}{1cm}|w{c}{1cm}|w{c}{1cm}|w{c}{1cm}|}
\hline
\multicolumn{8}{|c|}{$\gamma(\alpha, \beta, \mathrm{r}=4)$} \\
\hline
\diagbox[innerwidth=1.5cm]{$\beta$}{$\alpha$} & 1.4 & 1.5 & 1.6 & 1.7 & 1.8 & 1.9 & 2.0 \\
\hline 
0.40 & 2.43 & 2.47 & 2.50 & 2.53 & 2.56 & 2.58 & 2.60 \\ \hline
0.50 & 2.29 & 2.33 & 2.37 & 2.41 & 2.44 & 2.47 & 2.50 \\ \hline
0.60 & 2.14 & 2.20 & 2.25 & 2.29 & 2.33 & 2.37 & 2.40 \\ \hline
0.70 & 2.00 & 2.07 & 2.12 & 2.18 & 2.22 & 2.26 & 2.30 \\ \hline
0.80 & 1.86 & 1.93 & 2.00 & 2.06 & 2.11 & 2.16 & 2.20 \\ \hline
0.90 & 1.71 & 1.80 & 1.87 & 1.94 & 2.00 & 2.05 & 2.10 \\ \hline
1.00 & 1.57 & 1.67 & 1.75 & 1.82 & 1.89 & 1.95 & 2.00 \\ \hline
\end{tabular}
%
\centering
\caption{The spectral index $\gamma$ for a shock with compression ratio  $\mathrm{r}=3$}\label{tab:r3}
\begin{tabular}{|w{c}{2cm}|w{c}{1cm}|w{c}{1cm}|w{c}{1cm}|w{c}{1cm}|w{c}{1cm}|w{c}{1cm}|w{c}{1cm}|}
\hline
\multicolumn{8}{|c|}{$\gamma(\alpha, \beta, \mathrm{r}=3)$} \\
\hline
\diagbox[innerwidth=1.5cm]{$\beta$}{$\alpha$} & 1.4 & 1.5 & 1.6 & 1.7 & 1.8 & 1.9 & 2.0\\
\hline 
0.40 & 3.14 & 3.20 & 3.25 & 3.29 & 3.33 & 3.37 & 3.40 \\ \hline
0.50 & 2.93 & 3.00 & 3.06 & 3.12 & 3.17 & 3.21 & 3.25 \\ \hline
0.60 & 2.71 & 2.80 & 2.88 & 2.94 & 3.00 & 3.05 & 3.10 \\ \hline
0.70 & 2.50 & 2.60 & 2.69 & 2.76 & 2.83 & 2.89 & 2.95 \\ \hline
0.80 & 2.29 & 2.40 & 2.50 & 2.59 & 2.67 & 2.74 & 2.80 \\ \hline
0.90 & 2.07 & 2.20 & 2.31 & 2.41 & 2.50 & 2.58 & 2.65 \\ \hline
1.00 & 1.86 & 2.00 & 2.13 & 2.24 & 2.33 & 2.42 & 2.50 \\ \hline
\end{tabular}
\end{table}
\end{center}

\section{Conclusions}

In this work the theory of diffusive shock acceleration is extended to the case of non-classical transport with
L\'{e}vy flights and L\'{e}vy traps, when the mean square displacement grows nonlinearly with time. In this approach the Green function is not a Gaussian but it exhibits power-law tails. 

By using the propagator appropriate for non-classical diffusion, it is found for the first time that energy spectral index of particles accelerated at shock fronts is 
\begin{equation}\label{C6}
\gamma = \frac{\alpha(\mathrm{r} + 5) - 6 \beta}{\alpha(\mathrm{r} -1)},
\end{equation}
where $0 < \alpha \leq 2$ and $0 < \beta \leq 1$
are the exponents of power-law behavior of L\'{e}vy flights and L\'{e}vy traps, respectively.

We note that this result coincides with standard slope at $\alpha=2, \beta=1$ (normal diffusion), and also includes those obtained earlier for the subdiffusion ($\alpha=2, \beta < 1$)~\cite{Kirk:1996} and superdiffusion 
($1 < \alpha < 2, \beta=1$)~\cite{Perri:2012} regimes. When comparing~\eqref{C6} with result~\cite{Perri:2012}, it should be taken into account that the exponent $\mu$ of work~\cite{Perri:2012} is related to $\alpha$ by the equality $\mu = \alpha + 1$.

\section*{Funding}

The work is supported by the Russian Science Foundation (grant no. 23-72-00057). 

\section*{Appendix. Stable distributions}

Integral representations and the ones in the form of asymptotic series of one-dimensional symmetric and one-sided stable distributions are presented (see~\cite{Uchaikin:1999a}).

A1. One-dimensional symmetric stable distribution  $g_1^{(\alpha)}$.

\begin{equation*}
  g_1^{(\alpha)}(r)=\frac{\alpha r^{1/(\alpha-1)}}{\pi|1-\alpha|}\int_{0}^{\pi/2}{\exp\left[-r^{\alpha/(\alpha-1)}U(\varphi,\alpha,0)\right] U(\varphi,\alpha,0)d\varphi},
\end{equation*}
where
\begin{equation}\label{eq:u}
U(\phi,\alpha,\alpha')=\left [\frac{\sin(\alpha \phi + \alpha' \pi/2)}{\cos \phi}\right ]^{\alpha/(1-\alpha)}
\frac{\cos((\alpha-1)\phi + \alpha' \pi/2}{\cos \phi}.
\end{equation}

\begin{equation*}\label{eq:g1_r+}
  g_1^{(\alpha)}(r)=\frac{1}{\pi}\sum_{n=1}^{\infty}{\frac{(-1)^{n-1}}{n!}\Gamma\left(\frac{n}{\alpha}+1\right)\sin\left(n\frac{\pi}{2}\right)r^{n-1}},
\end{equation*}

\begin{equation*}\label{eq:g1_r-}
  g_1^{(\alpha)}(r)=\frac1\pi\sum_{n=1}^{\infty}{\frac{(-1)^{n-1}}{n!}\Gamma(n\alpha+1)\sin\left(n\alpha\frac{\pi}{2}\right)r^{-n\alpha-1}}.
\end{equation*}

A2. One-sided stable distribution $q_1^{(\beta)}(r)$.

\begin{equation*}\label{eq:q13}
q_1^{(\beta)}(r)=\frac{\beta r^{1/(\beta-1)}}{\pi|1-\beta|}\int\limits_{-\pi/2}^{\pi/2}{\exp\left[-r^{\beta/(\beta - 1)}U(\varphi,\beta,1)\right]U(\varphi,\beta,1)d\varphi},
\end{equation*}
\begin{equation*}\label{eq:q11}
q_1^{(\beta)}(r)=\frac{1}{\pi}\sum_{n=1}^{\infty}{\frac{(-1)^{n-1}}{n!}\Gamma\left(\frac{n}{\beta}+1\right)\sin\left(n\frac{\pi}{2}\right)r^{n-1}},
\end{equation*}
\begin{equation*}\label{eq:q12}
q_1^{(\beta)}(r)=\frac{1}{\pi}\sum_{n=1}^{\infty}{\frac{(-1)^{n-1}}{n!}\Gamma(n\beta+1)\sin(n\beta\pi)r^{-n\beta-1}},
\end{equation*}
where $U(\varphi,\beta,1)$ is defined by the equation~\eqref{eq:u}.

\end{document}